\documentstyle [aps,twocolumn,epsf]{revtex}
\catcode`\@=11 
\newcommand{\be}{\begin{equation}}
\newcommand{\ee}{\end{equation}}
\newcommand{\bea}{\begin{eqnarray}}
\newcommand{\eea}{\end{eqnarray}}

\newcommand{\p}{\partial}

\newcommand{\la}{\langle}
\newcommand{\ra}{\rangle}
\newcommand{\rd}{\mbox{d}}
\newcommand{\ri}{\mbox{i}}
\newcommand{\re}{\mbox{e}}

\begin {document}
\title{
Kagom{\'e} Lattice Antiferromagnet Stripped to Its Basics}

\author{ P. Azaria$~^{1}$,  C. Hooley$~^{2}$, P. Lecheminant$~^{3}$,
C. Lhuillier$~^{1}$,  and A. M.
Tsvelik$~^{2}$}

\vspace{0.5cm}

\address{$^1$
Laboratoire de Physique Th{\'e}orique des Liquides,\\
Universit{\'e} Pierre et Marie Curie, 4 Place Jussieu, 75252 Paris,
France\\
$^2$ Department of Physics, University of Oxford, 1 Keble Road,
Oxford OX1 3NP, UK\\
$^3$ Laboratoire de Physique Th{\'e}orique et Mod{\'e}lisation,\\
Universit{\'e} de Cergy-Pontoise, Site de Saint Martin,
2 avenue Adolphe Chauvin, 95302
Cergy-Pontoise Cedex, France}

\vspace{3cm}

\address{\rm (Received: )}
\address{\mbox{ }}
\address{\parbox{14cm}{\rm \mbox{ }\mbox{ }
 We study a model of a spin S = 1/2 Heisenberg antiferromagnet on
a one dimensional lattice with the local symmetry of the 
two dimensional kagom{\'e} lattice.
Using three complementary approaches, it is shown 
that the low energy spectrum can be described by two critical Ising models 
with different velocities. One of these velocities is small, leading to 
a strongly localized Majorana fermion.
These excitations are singlet ones  whereas 
the triplet sector has a spectral gap. 
}}
\address{\mbox{ }}
\address{\parbox{14cm}{\rm PACS No: 75.10.Jm, 75.40.Gb}}
\maketitle

\makeatletter
\global\@specialpagefalse
\makeatother

The famous kagom{\'e} lattice antiferromagnet still remains largely
a mystery after a decade of extensive studies.
This system exhibits both frustration and low 
coordinance and classically it 
has infinite continuous
degeneracies. Local distortions allow it
to explore its many ground states with no cost in energy
and lead to a very specific linear spin wave spectrum with
a whole branch of zero energy excitations\cite{spinwave}.
In the quantum case (S=1/2),
the system is likely 
to be a spin liquid with a gap for the magnetic 
excitations\cite{quantum}. 
For finite samples, the system
has a huge number of singlet states below the first
triplet \cite{phle} which is a rather unexpected feature for a
two-dimensional quantum antiferromagnet. Moreover,
the analysis of the specific heat shows the existence 
of unusual low-lying excitations\cite{eltsner}. 
The presence of these low-lying singlet excitations
below the spin gap gives
a picture of an intriguing spin liquid
that deserves understanding from a general 
point of view.

\begin{figure}
\begin{center}
\unitlength=1mm
\begin{picture}(80,60)(9,-10)
\put(15,10){\line(1,0){70}}
\put(15,40){\line(1,0){70}}
\multiput(20,10)(20,0){4}{\circle*{2}}
\multiput(20,40)(20,0){4}{\circle*{2}}
\multiput(30,25)(40,0){2}{\circle*{2}}
\multiput(20,10)(40,0){2}{\line(2,3){20}}
\multiput(20,40)(40,0){2}{\line(2,-3){20}}
\put(15,0){\vector(1,0){70}}
\multiput(20,0)(20,0){4}{\line(0,-1){2}}
\put(20,-5){\makebox(0,0){\small{$2j$}}}
\put(40,-5){\makebox(0,0){\small{$2j+1$}}}
\put(60,-5){\makebox(0,0){\small{$2(j+1)$}}}
\put(80,-5){\makebox(0,0){\small{$2(j+1)+1$}}}
\put(23,30){\makebox(0,0){$J_{\bot}$}}
\put(30,44){\makebox(0,0){$J_{\|}$}}
\end{picture}
\end{center}
\caption{One dimensional version of the kagom{\'e} lattice.}
\label{model1}
\end{figure}
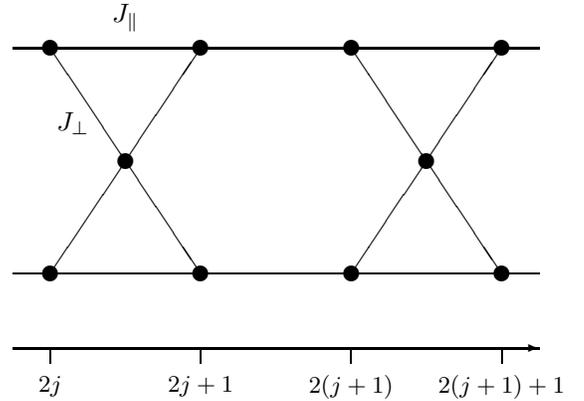

In this paper, we shall study a one-dimensional 
model of spin S = 1/2 Heisenberg antiferromagnet on a lattice 
presented on Fig.~\ref{model1} which retains the local symmetry of
the kagom{\'e} net. Insight
into the behavior of
the two-dimensional kagom{\'e}
antiferromagnet might be gained by investigating this simplified
system. In particular,
the one dimensional model may help us to identify the slow
degrees of freedom of the problem. 

According to the conventional wisdom the model we study may be viewed as 
a version  of a three-leg spin ladder which  in the low-energy limit 
one should expect to fall into the universality class of the S = 1/2
Heisenberg spin chain. The latter means 
that the low-lying excitations (often called spinons) are
represented by
one gapless bosonic mode (in the language of the theory of 
critical phenomena this means that the central charge of the model 
is equal to  C = 1 \cite{sierra}). However,
the frustration may play its tricks and, as we shall see, a somewhat
different scenario is realized. Namely, the bosonic mode decouples into 
two modes of real (Majorana) fermions  (C= 1/2 each) having  
different spectra.

The Hamiltonian for the model shown in Fig.~\ref{model1} may be
written
\bea
{\cal H} = \sum_{a=1,2} \sum_j [J_{\|}\left(
{\bf S}_{a,2j} \cdot {\bf
S}_{a,2j+1} + {\bf S}_{a,2j+1} \cdot {\bf 
S}_{a,2j+2} \right)  \nonumber\\  
+ J_{\bot} {\bf s}_{2j+{1
\over 2}} \cdot \left( {\bf S}_{a,2j} + {\bf S}_{a,2j+1}\right)]  
\label{oldint}
\eea
where ${\bf s}_{2j+{1 \over 2}}$ are the ``middle spins'', and
${\bf S}_{a,j}$ are the ``rail spins'', the index $a$ taking two
values ($a=1,2$), one for each rail.
All interactions are antiferromagnetic
and 
we shall also consider the case where 
interaction  between spins belonging to the rails of the
lattice is much stronger than the interactions with the middle spins: 
$J_{\|} >> J_{\perp}$. Under this condition, one can 
describe the
chains in the continuous approximation, representing the spins on 
each chain as a sum
of ferromagnetic (${\bf M}_a$) and staggered (${\bf n}_a$) parts:
$
{\bf S}_{a,j} \rightarrow {\bf M}_a (x) + (-1)^j{\bf n}_a (x). 
$
It is crucial for our analysis that the middle spins 
interact only with the ferromagnetic part of the magnetization. 
The latter has the following remarkable property - it can be 
written as a sum of ``currents''
$
{\bf M}_a = {\bf J}_a + \bar{\bf J}_a, 
$
where the currents satisfy the 
same commutation relations as 
 bilinear combinations of left- and right-moving Dirac fermions\cite{affleck}:
\bea
\bar {\bf J}_a = R_{a,\alpha}^+\vec\sigma_{\alpha\beta}R_{a,\beta}/2, 
{\bf J}_a = L_{a,\alpha}^+\vec\sigma_{\alpha\beta} L_{a,\beta}/2
\eea
where $\vec\sigma$ are the Pauli matrices.
The corresponding algebra (SU(2)$_1$) 
is called the level $k = 1$ SU(2) Kac-Moody (KM) algebra 
where $k$ refers to the number of species of spin-1/2 fermions (one in the 
given case). Now we notice that  the interaction includes 
 the sum of two SU(2)$_1$ spin currents
${\bf J} = {\bf J}_1 + {\bf J}_2$ which, by definition, is  a 
level $k = 2$ current. These currents have the same commutation relations as 
currents of three Majorana 
fermions (the explicit expression is given later) \cite{zamolo,shelton}.
Since each Dirac fermion can be 
represented as a linear 
combination of two Majorana (real) fermions, it 
is clear that only 3/2 of the low-energy  degrees of 
freedom of rails are involved in the 
interaction. Obviously the other decoupled degrees 
of freedom remain critical. 
Using the results of Ref. \cite{shelton} we can represent 
 the continuous limit of the Hamiltonian (\ref{oldint})
as follows: ${\cal H} = {\cal H}_s + {\cal H}_t$ with
\bea
{\cal H}_s = - \frac{\ri v}{2}
\int \rd x \; \left(r_0\p_xr_0 - \; l_0\p_xl_0\right) 
\label{sing}\\ 
{\cal H}_t =  
\frac{\pi v}{2}\int \rd x \;[{\bf J}^2(x) 
+ \bar{\bf J}^2(x)] \nonumber \\ 
+ \sum_j g \int \rd x \;\delta(x - aj)\; {\bf s}_j
\cdot [{\bf
J}(x) + \bar{\bf J}(x)].
\label{ham1}
\eea
Here $v \sim J_{\|}$ is the spin velocity which
we shall later put to unity and  the
coupling constant is $g \sim J_{\perp}$.
The Majorana fermion ($r_0, l_0$) describes 
non-magnetic, gapless excitation. It 
represents Ising degrees of freedom that 
do not interact with the central spins and are associated with 
a discrete $Z_2$ interchange symmetry between 
the surface chains.
We shall study the Hamiltonian (\ref{ham1}) using several 
alternative approaches. First, we shall study an integrable deformation 
of model (\ref{ham1}) . To make sure that the exact solution 
reproduces the qualitative features of the spectrum, 
 we shall consider an anisotropic version of (\ref{ham1}) bearing in mind 
that it may simplify in the limit of strong 
anisotropy  (the so-called Emery-Kivelson limit\cite{emery}).
Finally, a direct mean field treatment of the 
lattice in the isotropic limit will be performed. 

{\bf Exact solution of a related model}.
Since the currents satisfying the level $k=2$ 
SU(2) KM algebra can be represented as
fermionic bilinears, 
the effective theory for the central spins corresponds to 
a special version of the two-channel ($k = 2$!) Kondo-lattice model where 
 ``electrons'' do not experience backscattering. 
A similar model (with $k = 1$) was
considered in Ref.\cite{sak}; the result is that
integrability is achieved if one adds an additional interaction:
\be
{\cal H}_{ex} = {\cal H}_{t}
+ g'\int \rd x \; {\bf J}(x)\cdot \bar{\bf J}(x) \label{integ}
\ee
where the ratio $g'/g$ is fixed. 
The Bethe ansatz equations for model
(\ref{integ}) are derived in the same way as in Ref. \cite{sak} with the
only difference that ``conduction electrons'' now belong to the spin-1
representation of the SU(2) group. 
The result is 
\be
[e_2(x_a + 1/g)e_2(x_a - 1/g)]^N e_1^{N_0}(x_a) = \prod_{b = 1}^Me_2(x_a
- x_b)
\ee
where 
$2N$ and $N_0$ are the numbers of spins on the
rails and in the middle, respectively, $M$ is the number of up spins
($M = 2N + N_0/2 - S^z$) and
$
e_n(x) = (x - \ri n/2)/(x + \ri n/2).
$
The total energy takes the form
\be
E = \frac{1}{2\ri}\sum_a\ln[e_2(x_a + 1/g)/e_2(x_a - 1/g)].
\ee
Following the standard procedures of the Bethe ansatz, we derive the
following system of integral equations for the free energy:
\bea
F &=& - N T\int \rd x \;[s(x + 1/g) + 
s(x - 1/g)]\ln\left[1 +
 \re^{\epsilon_2(x)/T}\right] \nonumber \\ 
&-& N_0 T \int \rd x \; s(x)\ln\left[1 +
 \re^{\epsilon_1(x)/T}\right], \label{free}
\eea
with
\bea
\epsilon_j(x) = Ts*\ln\left[1 + \re^{\epsilon_{j -
 1}(x)/T}\right]\left[1 + \re^{\epsilon_{j + 1}(x)/T}\right] \nonumber \\
-
 \delta_{j,2}\Delta \cosh\pi x
\label{epsrec}
\eea
and $\lim_{n \rightarrow \infty}\epsilon_n(x)/n = H$,
$H$ being the magnetic field. In (\ref{epsrec}), one has
$\Delta \sim \exp(- \pi/g)$, 
$s(x) = [2\cosh\pi x]^{-1}$, and the convolution 
product is denoted by $*$. 

 At low temperatures $T << \Delta$, one has 
$
\epsilon_2(x) \approx - \Delta \cosh\pi x, ~~ \epsilon_1(x) = 
O(\re^{-\Delta/T}).
$
In this case the first term in the free energy (\ref{free}) gives an
exponentially small contribution corresponding to excitations with a
spectral gap $\Delta$ and the last term gives
$
F \approx - TN_0\ln\sqrt 2
$
showing that each central spin contributes $\ln\sqrt 2$ to the ground
state entropy.
A  lesson we learn from the exact solution is that there are
degrees of freedom presumably localized on central spins 
(and their number corresponds exactly to a single
Majorana mode) which remain decoupled. One may conjecture that since
the magnetic modes are separated from the ground state by a gap,
the soft modes will remain soft even when one departs from the
integrable point. Below we shall give more rigorous arguments to
support this conjecture. 

{\bf The Emery-Kivelson limit}.
We shall now study a U(1) version of the Hamiltonian (\ref{ham1}),
characterized by anisotropic interactions 
with $g\rightarrow g_{\parallel}, g_{\perp}$
where in the limit of strong anisotropy, the 
low energy degrees of freedom can be identified.
This approach has been  fruitfully applied
in quantum impurity problems and gives a 
simple description of the two-channel Kondo model \cite{emery}
and also of the Kondo lattice \cite{zachar}. 
In particular, for the two-channel Kondo problem,
Emery-Kivelson identified the residual zero 
point entropy stemming from a unique Majorana zero mode. 
To reformulate the Hamiltonian (\ref{ham1}) as 
a fermionic theory similar to Kondo 
problems, we first introduce   
three right- and left-moving Majorana 
fermions $r_b, l_b$ ($b=1,2,3$)
to use the fermionic 
representation of  
the SU(2)$_2$ spin current:  
${\bar J}^a =  - \ri \epsilon^{abc} r_b r_c /2$
with a similar relation for the left-moving current \cite{zamolo}. 
The next step in our solution is to combine the two Majorana
fermions ($r_1,l_1$) and ($r_2,l_2$) to form a single
Dirac spinor ($R,L$) which in turn can be bosonized. Introducing
a massless bosonic field $\Phi$ and its dual
field $\Theta$, one can write:
\bea
R = \frac{r_1 + \ri r_2}{\sqrt 2} =
\frac{\epsilon}{\sqrt{2\pi a_0}}\exp[\ri\sqrt\pi(\Phi -
\Theta)]\nonumber\\
L = \frac{l_1 +\ri l_2}{\sqrt 2} =
\frac{\epsilon}{\sqrt{2\pi a_0}}\exp[-\ri\sqrt\pi(\Phi +
\Theta)] \label{dirac}
\eea
where $a_0$ is a short distance cut-off.
The anticommutation relation between $R$ and $L$ is taken into account by the
commutator: $[\Theta(y), \Phi(x)] = -\ri \theta(y -x)$.
The real fermionic zero mode ($\epsilon$)
is necessary to establish
the correct anticommutation relations with
the third Majorana fermion ($r_3,l_3$).
The interacting part of the Hamiltonian (\ref{ham1})
is then given by:
\bea
{\cal H}_{int} = \sum_j \left(
-\frac{g_{\parallel}}{\sqrt\pi}s^z_{j}(\p_x\Phi)_{j}
+ 
\frac{g_{\perp}}{\sqrt{4\pi
a_0}}\epsilon_j s^+_{j}\re^{i\sqrt\pi\Theta_{j}} \right. \nonumber \\
\left.
\left(\re^{i\sqrt{\pi}\Phi_{j}}\re^{-i\pi/4} l_{3,j}
+\re^{-i\sqrt{\pi}\Phi_{j}}\re^{i\pi/4} r_{3,j}\right)
+ h.c \right).
\eea
Following Emery and Kivelson, we absorb the phase factor
$\re^{\pm i\sqrt{\pi} \Theta_{j}}$ into the spin operators by a unitary
transformation $U$:
\be
U = \re^{-i\sqrt{\pi} \sum_j\Theta_{j}s^z_{j}}.
\ee
As a result the interaction becomes
\bea
{\cal H}_{int} \rightarrow U {\cal H}_{int} U^{\dagger}
= \sum_j \left( -\frac{(g_{\parallel} -
\pi)}{\sqrt\pi} s^z_{j}(\p_x\Phi)_{j} \right. \nonumber\\
\left. 
+
\frac{g_{\perp}}{\sqrt{4\pi
a_0}}a^+_{j}\left(\re^{i\sqrt{\pi}\Phi_{j}}l_{3,j}
+(-1)^{j}\re^{-i\sqrt{\pi}\Phi_{j}}r_{3,j}\right) + h.c.
\right)
\label{intr}
\eea
where we have replaced the combinations of spin-1/2 operators and
fermionic zero mode by local fermions ($a_{j}$) using the Jordan-Wigner
transformation. We have also absorbed a phase factor $(-\ri)^{j}$ 
in the definition of $a^+_{j}$. 

At a special point 
$g_{\parallel} = \pi$ (called the Toulouse point), part of the 
interaction vanishes and the low energy physics  
can be studied by a simple mean-field theory. 
Since the scaling dimension 
$d$ of
the bosonic exponents
in (\ref{intr}) 
is pretty small ($d = 1/4$), we shall
replace them by their averages and try to
solve the problem self-consistently. Introducing two Majorana
fermions ($\xi_{1,2}$) associated with the complex fermion $a$:
$a = (\xi_1 + \ri\xi_2)/\sqrt 2$, 
we find that the effective action decouples 
into bosonic and fermionic parts in the mean field
limit: ${\cal S}_{MF} =
{\cal S}_{B}+ {\cal S}_{F}$ with
\bea
{\cal S}_{B} = \int dx d\tau \; [
\frac{1}{2}\left(\partial_{\mu} \Phi\right)^2
+ \lambda \cos\left(\sqrt \pi \Phi\right)] \nonumber \\
{\cal S}_{F} = \sum_j \int d\tau \;[
\frac{1}{2}\xi_{1,j} \partial_{\tau} \xi_{1,j} +
\frac{1}{2}\xi_{2,j} \partial_{\tau} \xi_{2,j} \nonumber\\
+
\frac{1}{2}\chi_{j} \partial_{\tau} \chi_{j} 
- \ri \chi_{j+1} \chi_{j} - \ri \Delta \xi_{2,j} \chi_{j}]  
\label{meanfield}
\eea
where $\chi$ is a Majorana fermion express in terms 
of the right- and left-moving Majorana fields ($r_3,l_3$):
$\chi_{j} = (l_{3,j} + (-1)^j r_{3,j})/\sqrt{2}$. 
The mean-field parameters are defined by: $\lambda = 
-\ri g_{\perp} \langle \xi_2 \chi\rangle/\sqrt{\pi a_0}$
and $\Delta = g_{\perp} \langle 
\cos(\sqrt{\pi} \Phi)\rangle/\sqrt{\pi a_0}$. 
Solving the self-consistency equations, we find that
the bosonic field $\Phi$ becomes massive with 
a gap of the order 
$\Delta \sim g_{\perp}^{4/3}\ln^{1/6}(1/ g_{\perp}^2a_0)$. 
In the fermionic sector, the Majorana fermion $\xi_2$
hybridizes with the Majorana field $\chi$ which 
in terms of the original spins stems from the rails spins. 
The resulting excitation spectrum is reminiscent of the 
one found in Kondo lattices\cite{alexei} with a 
small gap $\Delta_g \sim \Delta^2/J_{\parallel} \ll \Delta$. 
Finally, there is still a singlet localized Majorana 
fermion $\xi_1$ which decouples from the conduction
sea at the Toulouse point. This mode gives a 
zero point entropy of magnitude ${1 \over 2} \ln 2$ per central
spin as found above in the integrable model.
Away from the Toulouse point ($\delta g_{\parallel} = 
g_{\parallel} - \pi \ne 0$), this mode will acquire
a small dispersion and will contribute to the specific
heat. Coming back to the original model, adding the 
contribution of the singlet Majorana fermion $(r_0,l_0)$
which has decoupled from the interaction,
the total central charge in the long-distance limit is
C=1. Let us stress that the two Majorana modes, in the 
singlet sector, contributing
to the central charge are of different {\it nature}. The Majorana
field $(r_0,l_0)$ describes a critical Ising model whereas 
$\xi_1$ is a strongly localized Majorana fermion.
In contrast, the triplet sector  has a small spectral gap.
As a result of this spectrum, the middle spins are
disordered and have short-ranged spin correlations.

The experience gained from the study of Kondo models leads
us to expect that the preceeding results obtained in the
Emery-Kivelson limit of the model will extend
to the isotropic point. To show this,
we shall now develop a mean-field theory
directly at the isotropic limit
keeping track of the lattice structure more
accurately than in the previous approach.
 
{\bf Direct mean-field approach}.
The Hamiltonian (\ref{ham1}) is reformulated
in terms of a fermionic model on the lattice: 
\bea
{\cal H}_t = \ri J_{\parallel}\sum_j\chi_{j+ 1}^a\chi_{j}^a -
\frac{ig}{2}\sum_j{\bf s}_{2j
+ {1 \over 2}}\; . \;\left([\vec \chi_{2j}\times\vec \chi_{2j}]
\right. \nonumber\\
+\left. [\vec
\chi_{2j + 1}\times\vec \chi_{2j + 1}]\right). \label{fermham}
\eea
In the continuum limit, the rail of $\chi^a$ fermions 
will contain left- and right- moving Majorana fermions
($r_a, l_a; a=1,2,3$) of the SU(2)$_2$ spin current: 
$ 
\chi_j^a = 
-l_a\left(x\right) - \left(-1\right)^j r_a
\left(x\right);
$
with this identification, the Hamiltonian (\ref{ham1})
is then reproduced.
The model (\ref{fermham}) corresponds to local moments (${\bf s}_{2j+1/2}$)
interacting with a sea of three Majorana fermions ($\chi_{j}^a$). 
To describe the middle spins
we use the Majorana representation for spins S=1/2  
(see Ref. \cite{alexei} and references therein):
\be
{\bf s}_n = -\frac{\ri}{2}[\vec \gamma_{n}\times\vec \gamma_n] ,
\quad n = 2k+{1 \over 2}, k \in {\cal Z} \label{locma} 
\ee
where $\gamma^a_n$ are local Majorana fermions satisfying
the anticommutation relations 
$
\{\gamma^a_n,\gamma_m^b\} = \delta_{nm}\delta^{ab}.
$
This representation
(\ref{locma}) reproduces the spin commutation relations and gives the
correct value of the Casimir operator: ${\bf s}_n^2=3/4$.

\begin{figure}
\begin{center}
\unitlength=1mm
\begin{picture}(100,30)(6,0)
\put(10,10){\line(1,0){80}}
\multiput(20,10)(40,0){2}{\line(1,1){10}}
\multiput(40,10)(40,0){2}{\line(-1,1){10}}
\multiput(20,10)(20,0){4}{\circle*{2}}
\multiput(30,20)(40,0){2}{\circle*{2}}
\put(20,6){\makebox(0,0){$\chi_1$}}
\put(40,6){\makebox(0,0){$\chi_2$}}
\put(60,6){\makebox(0,0){$\chi_3$}}
\put(80,6){\makebox(0,0){$\chi_4$}}
\put(30,24){\makebox(0,0){$\gamma_1$}}
\put(70,24){\makebox(0,0){$\gamma_2$}}
\put(20,16){\makebox(0,0){$U_-$}}
\put(40,16){\makebox(0,0){$U_+$}}
\put(60,16){\makebox(0,0){$V_-$}}
\put(80,16){\makebox(0,0){$V_+$}}
\end{picture}
\end{center}
\caption{The unit cell in the mean-field approximation.}
\label{unitcell}
\end{figure}
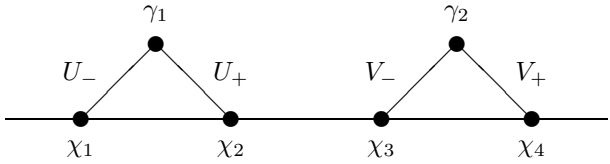

A  model very similar to (\ref{fermham}) was analyzed in Ref.
\cite{coleman}. Following this analysis 
we substitute (\ref{locma}) into (\ref{fermham}) and decouple the
interaction with an auxiliary field $V$ living on the links connecting
rails with middle spins.
In the mean-field
approximation the variables $V$ are considered as static with their
values determined self-consistently by minimizing the free energy. 
 We find that the minimum is achieved when a  
unit cell contains two middle spins (see Fig.~\ref{unitcell})
with either $U_+ = V_- = 0, U_- = \pm V_+
\equiv \Delta$ or $U_- = V_+ = 0, U_+ = \pm V_- \equiv \Delta$.
All  branches of the
spectrum are found to be gapful.
The low energy band is rather flat towards the edge of the
Brillouin zone in agreement with the Emery-Kivelson limit
of the model in the triplet sector.
However, as 
was demonstrated in Ref.\cite{coleman}, the local Z$_2$-degeneracy of
the ground state generates a local real fermionic zero mode $\gamma_0$
which is coupled to the three Majorana band fermions with
the Lagrangian:
\be
L = \sum_j[\frac{1}{2}\gamma^0_j\p_{\tau}\gamma^0_j 
+ g '\gamma^0_j\chi^1_j\chi^2_j\chi^3_j].
\ee
This singlet zero mode 
acquires a small dispersion with the bandwidth 
$\Delta_g \sim \Delta^2/J_{\parallel}\ll \Delta$. 
The total amount of entropy accumulated in this band 
is ${1 \over 2} \ln 2$ per central spin and this degeneracy 
will be slightly lifted and results in the coherent
dispersion of this mode. This mode corresponds to the 
strongly localized Majorana fermion found in the Emery-Kivelson limit.
This field will manifest itself in the 
spin-spin correlation functions of the central spins since 
at low energies, they behave like
$
{\bf s} \sim \gamma_0\vec \chi.
$
Due to the small dispersion of the fermionic mode $\gamma_0$,
correlation functions of
middle spins are strongly 
localized in space, but not in time where the
characteristic scale is $\sim \Delta_g^{-1}$:
\be
\la\la{\bf s}(\tau,j){\bf s}(0,j)\ra\ra \sim K_1(\Delta_g|\tau|)
\ee
which for small times is proportional to $1/|\tau|$, like for the 
two-channel Kondo problem.
  
In conclusion, the  version of the kagom{\'e} 
lattice studied in this paper can be reformulated
as a fermionic theory similar to models of Kondo lattices. 
Its low-energy excitations are 
two spin-singlet Majorana modes with 
 different spectra: a critical 
Ising mode and a strongly localized Majorana fermion, 
whereas
the triplet sector has a small spectral gap. The physics of the 
localized low-energy mode is similar to physics of the two-channel Kondo model. 
This picture
retains some properties of the kagom{\'e}
antiferromagnet with very soft singlet modes and a gap for
magnetic excitations. It is tempting
to conjecture that the singlet degrees of freedom
and the gapful magnetic excitation identified here might be responsible 
for the additional structure seen in the specific
heat of the kagom{\'e} magnet at low temperature.  

A.~M.~Tsvelik acknowledges the kind hospitality of Ecole Normale
Sup{\'e}rieure during his stay in Paris.
The authors thank P.~Chandra, P.~Coleman, B.~Dou{\c c}ot, 
H.-U. Everts, and Ch.~Waldtmann for important discussions.


\begin{thebibliography}{99}
\bibitem{spinwave} J. T. Chalker, P. C. W. Holdsworth, and
E. S. Shender, Phys. Rev. Lett. {\bf 68}, 855 (1992);
A. B. Harris, C. Kallin, and
A. J. Berlinsky, Phys. Rev. B {\bf 45}, 2899 (1992); 
A. Chubukov, Phys. Rev. Lett. {\bf 69}, 832 (1992);
P. Chandra, P. Coleman, and I. Ritchey,
J. Phys. I France {\bf 3}, 591 (1993).
\bibitem{quantum} C. Zeng and V. Elser, Phys. Rev. B {\bf 42},  
8436  (1990);
J. T. Chalker and J. F. G. Eastmond, Phys.
Rev. B {\bf 46},  14201  (1992); R. R. P. Singh and D. A. Huse,
Phys. Rev. Lett. {\bf 68}, 1766 (1992); 
S. Sachdev, Phys. Rev. B {\bf 45}, 12 377 (1992);
P. W. Leung and V. Elser, Phys. Rev. B {\bf 47},
5459 (1993).
\bibitem{phle} P. Lecheminant, B. Bernu, C. Lhuillier, L. Pierre, and
P. Sindzingre, Phys. Rev. B {\bf 56}, 2521 (1997).
\bibitem{eltsner} N. Eltsner and A.P. Young, Phys.
Rev. B {\bf 50},  6871  (1994).
\bibitem{sierra} G. Sierra, J. Phys. A {\bf 29}, 3299 (1996).
\bibitem{affleck} I. Affleck, Nucl. Phys. B{\bf 265}, 409 (1986). 
\bibitem{zamolo}  A. B. Zamolodchikov and. V. A.
Fateev, Sov. J. Nucl. Phys. {\bf 43}, 657 (1986).
\bibitem{shelton} D. G. Shelton, A. A. Nersesyan, and A. M. Tsvelik,
Phys. Rev. B {\bf 53}, 8521 (1996);
D. Allen and D. S{\'e}n{\'e}chal,
Phys. Rev. B {\bf 55}, 299 (1997).
\bibitem{emery} V. J. Emery and S. A. Kivelson, Phys. Rev. B {\bf 46},
10 812 (1992).
\bibitem{sak} E. H. Rezayi and J. Sak, Phys. Lett. {\bf 89A}, 451 (1982).
\bibitem{zachar} O. Zachar, S. A. Kivelson, and V. J. Emery,
Phys. Rev. Lett. {\bf 77}, 1342 (1996);
P. Coleman, A. Georges, and A. M. Tsvelik,
J. Phys. Cond. Matt. {\bf 9}, 345 (1997).
\bibitem{alexei} A. M. Tsvelik, Phys. Rev. Lett. {\bf 69}, 2142 (1992); 
P. Coleman, E. Miranda, and A. M. Tsvelik, 
Phys. Rev. B {\bf 49}, 8955 (1994). 
\bibitem{coleman} 
P. Coleman, L. B. Ioffe, and A. M. Tsvelik, 
Phys. Rev. B {\bf 52}, 6611 (1995).
 
\end{thebibliography}
\end{document}